\RequirePackage{fix-cm}
\documentclass[smallextended]{svjour3}       % onecolumn (second format)
\usepackage{float}
\usepackage{comment}
\usepackage{amsmath}
\usepackage{xcolor}
\newcommand{\be}{\begin{equation}}
\newcommand{\ee}{\end{equation}}
\newcommand{\bc}{\begin{comment}}
\newcommand{\ec}{\end{comment}}

\smartqed  % flush right qed marks, e.g. at end of proof
\usepackage{graphicx}
\begin{document}
	
	\title{Polarization Effects of Electro-Optic Sampling and Over-Rotation for High Field THz Detection}
	%\subtitle{Do you have a subtitle?\\ If so, write it here}
	
	%\titlerunning{Short form of title}        % if too long for running head
	
	\author{Gregory Bell         \and
		Michael Hilke %etc.
	}
	
	%\authorrunning{Short form of author list} % if too long for running head
	
	\institute{Gregory Bell \at
		Department of Physics, McGill University, Montreal, Canada \\
		\email{gregory.bell@mail.mcgill.ca}           %  \\
		%             \emph{Present address:} of F. Author  %  if needed
		\and
		Michael Hilke \at
		Department of Physics, McGill University, Montreal, Canada
	}
	
	\date{Received: date / Accepted: date}
	% The correct dates will be entered by the editor

	\maketitle

\section*{Abstract}

With ever increasing availability of terahertz fields, it is important to find suitable detection techniques without compromising the measured dynamic range. Electro-optic terahertz sampling techniques, which are commonly used to detect terahertz fields, exhibit over-rotation at high fields that limit the detection accuracy. Here we discuss a method to correct for over-rotation that put no limits on measured terahertz field strengths, while preserving the low field sensitivity. We further evaluate the induced polarizations at high terahertz fields and show how over-rotation can be corrected by simultaneously measuring the polarizations before and after the quarter wave plate. 

\section*{Introduction}

Terahertz (THz) spectroscopy \cite{dexheimer2007terahertz}--including time-resolved THz spectroscopy (TRTS), terahertz time-domain spectroscopy (THz-TDS), and THz emission spectroscopy--is concerned with what frequencies of a THz field emanate from a sample of interest. Planken and co-workers have pioneered the use of electro-optic (EO) crystals for the detection of THz signals \cite{planken2001measurement}. However, with the advent of ever more powerful THz fields, effects such over-rotation \cite{ibrahim2016ultra} can limit the applicability of EO crystals. Moreover, the approximations used in Planken have to be modified for higher fields. This is particularly timely with the ever increasing THz fields. Over two decades ago, THz field levels at the focal point were already reaching values in the hundreds of kV/cm \cite{You:93,538792}. A decade later, these field levels were reached without focusing \cite{kim2007terahertz,van2007single,bartel2005generation}. Indeed, the use of high THz fields, $\geq$1 MV/cm \cite{hirori2011single}, has seen increased usage in recent years, with beautiful experiments on inducing insulator-to-metal transitions in a metamaterial \cite{liu2012terahertz} and electronic and magnetic excitations in a ferromagnet \cite{shalaby2016simultaneous}. Other experiments include high THz field induced superconductivity \cite{matsunaga2013higgs,matsunaga2014light}, THz field-induced ferroelectric phases \cite{li2019terahertz,nova2019metastable}, photocarrier dynamics in monolayer graphene \cite{razavipour2015high,Mousavian:18}, and damage to thin metal films \cite{PhysRevLett.120.085704}. The number of high field THz emitters are also increasing over time, such as from a thin foil interaction \cite{Woldegeorgis:18}, from metal wires \cite{doi:10.1063/1.5031873}, and from organic crystals \cite{vicario2015high}. Sources are reaching field strengths up to the 10's of MV/cm \cite{Vicario:14,Vicario:15,vicario2017subcycle}. 
Hence, properly characterizing high field THz polarization becomes paramount.

Laser probe beam polarization is at the heart of THz signal detection. The intensity difference in the horizontal and vertical components of the beam relay the electric field strength and direction of the THz signal. For instance, in a naive depiction of the polarization of the probe beam due to EO sampling, where an elliptical polarization is created after the quarter-wave plate (QWP) with the major or minor axis at \(45^{\circ}\) from the vertical \cite{lee2009principles,peiponen2012terahertz}, this would not show any sign of THz. This is based on the vertical and horizontal components of the probe beam being not equal, and an ellipse at 45\(^{\circ}\) has equal vertical and horizontal components. This illustrates the importance to evaluate the complete polarization behaviour and this for the full range of EO sampling, which we present below. 

\begin{figure}[H]
\centering
\includegraphics[width=2.5in]{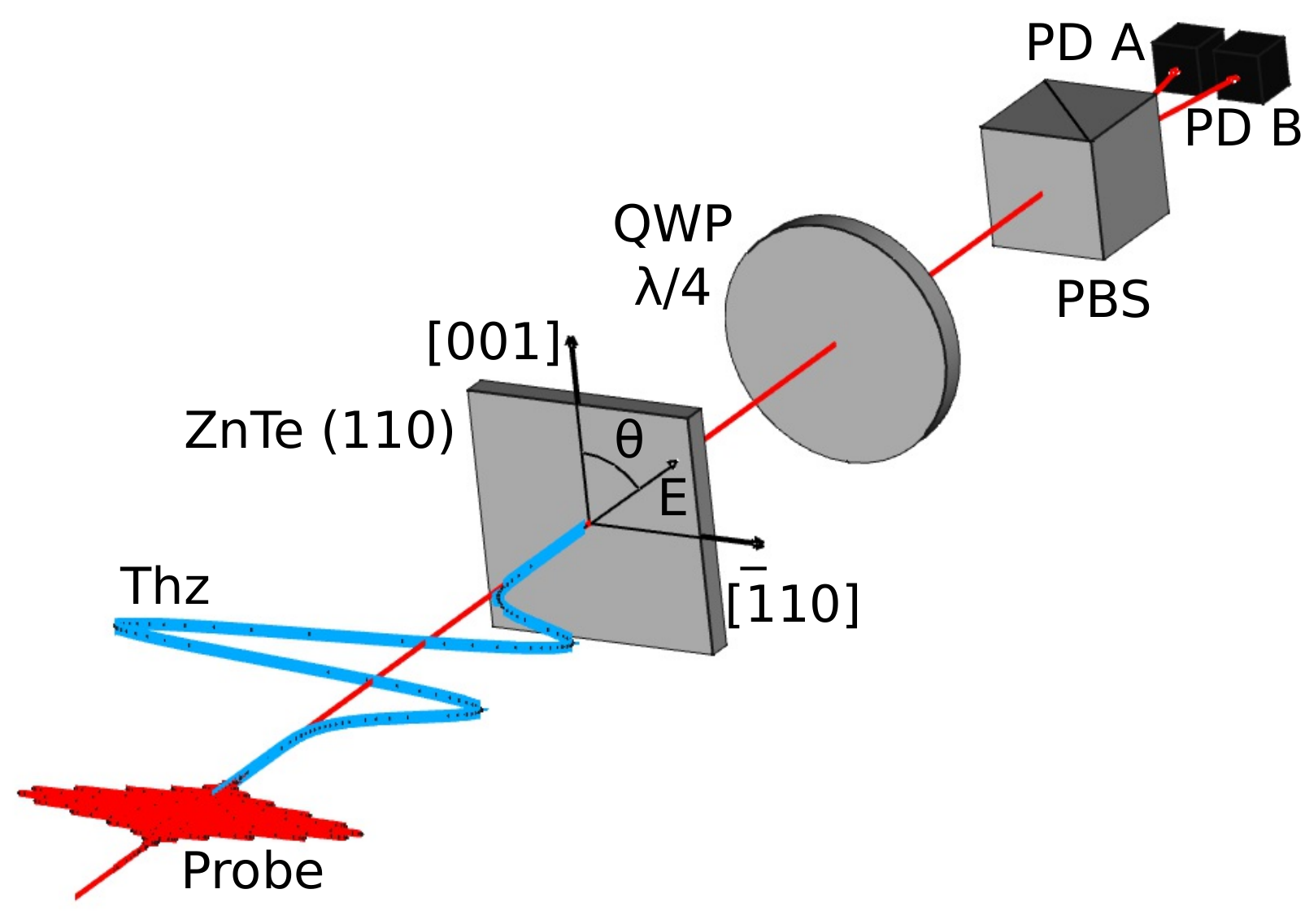}
\caption{EO sampling setup, with electric field (E) angles $\theta$. The THz meets with the probe beam in the ZnTe crystal, with the QWP used to adjust polarization of the probe, the PBS to split the vertical and horizontal components of the beam, and the PD's to detect the intensities of each polarization.}
\label{eosampznteaxes}
\end{figure}

The quasi-static THz electric field creates a changing waveplate in the EO crystal for the laser probe beam pulses. The phenomenon of creating a controllable waveplate in an EO crystal with an electric field due to non-linear optics has been known for decades \cite{yariv1989quantum}. The indexes of refraction are altered in different directions, causing the polarization of the beam to change. If a linearly polarized THz wave enters a EO crystal at \(\theta=90^{\circ}\) (Angle $\theta$ is shown in Figure \ref{eosampznteaxes}.), and the indexes or refraction are altered so their difference in the horizontal \(\theta=90^{\circ}\) and vertical \(\theta=0^{\circ}\) directions is increased, this would have no effect on the polarization--only the speed of the beam may change. To get the maximum change in polarization, the fast axis of the waveplate would have to be at \(\theta=45^{\circ}\). This angle corresponds to the angle that the fast axis of the EO crystal is at (after taking axis rotations into account as described in ref. \cite{planken2001measurement}), which the THz field creates when it also is horizontally linearly polarized. This may seem counter-intuitive with both probe beam and THz beam linearly polarized in the same direction. However, lack of symmetries in the crystals make this fast axis at \(\theta=45^{\circ}\) created by the THz possible.

In this work we present a detailed analysis of the changes in polarization in the probe beam in the presence of the THz field and the EO crystal at all THz field intensities. We discuss how the information of the polarization before and after the quarter wave plate (QWP) can be used to correct for issues such as over-rotation without decreasing the dynamic range. 

\begin{figure}[H]
    \centering
    \includegraphics[width=3in]{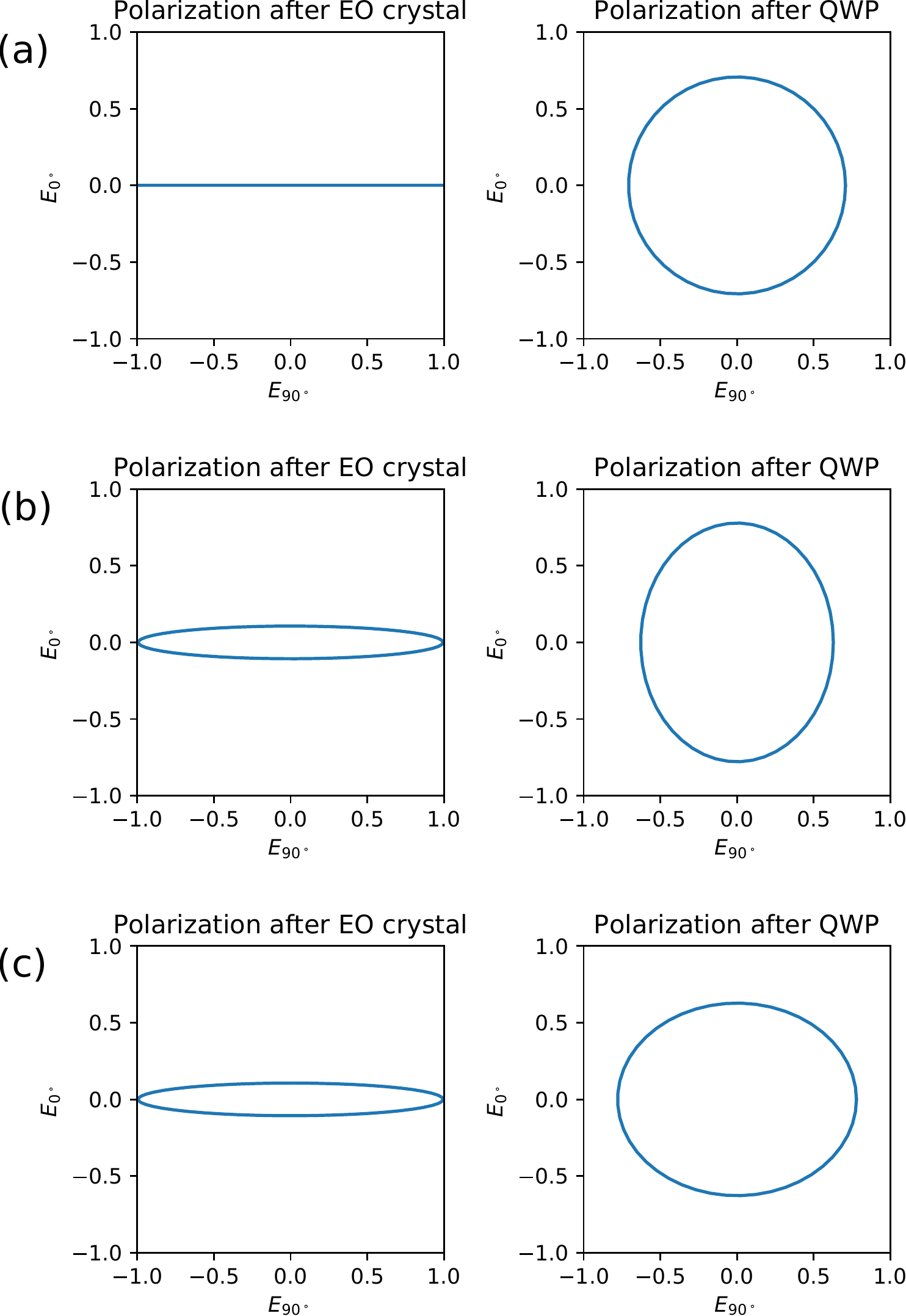}
    \caption{(a) Polarizations (using electric field directions, $E_{0^{\circ}}$ and $E_{90^{\circ}}$) of the probe beam, without THz. (b)  Polarizations with THz causing a phase change wave plate in the EO crystal. Polarization simulations are considered with a (110) oriented ZnTe crystal that is 1 mm thick (along the direction of propagation of the THz and probe beams). Here $E_{THz}=3~kV/cm$ is used. Plotting is explained in the discussion section. (c) Polarizations with THz causing a negative phase change wave plate in the EO crystal, with $E_{THz}=-3~kV/cm$. Polarization depictions used here were first shown (partially) in \cite{bell2017terahertz}}. 
    \label{fig:eop03kvpcm}
\end{figure}

Figure \ref{eosampznteaxes} shows a typical EO sampling setup, including an EO crystal (commonly ZnTe) where the probe and THz beams meet, a QWP, a Wollaston prism (a polarizing beamsplitter--PBS), and two balanced photodiodes (PD A and PD B) \cite{lee2009principles}. Here we assume the probe beam and THz pulse are p-polarized (horizontal in the lab frame if the beam stays at the same height level). The reason the QWP is needed is so the THz field can be mapped out with the probe beam for ``positive" and ``negative" directions (the distinctions are arbitrary and can be considered to correspond to ``left" and ``right" in the horizontal plane). With no THz, the probe continues to be linearly polarized after the EO crystal, and becomes circularly polarized after the QWP, as seen in Figure \ref{fig:eop03kvpcm}(a). 

When THz is present, the probe beam becomes elliptical before and after the QWP, as seen in Figure \ref{fig:eop03kvpcm}(b). However, when the THz wave (now considered negative) causes an equal magnitude phase change in the opposite direction, as seen in Figure \ref{fig:eop03kvpcm}(c), the polarization before the EO crystal looks identical to the previous phase change. After the QWP, the horizontal and vertical intensities \((I\propto E^{2})\) are different with these different phase changes. Here we can consider \(E_{0^{\circ}}>E_{90^{\circ}}\) as positive, and \(E_{90^{\circ}}>E_{0^{\circ}}\) as negative, representing opposite directions of THz field.

\section*{Methods}

\begin{figure}[H]
    \centering
\includegraphics[width=3in]{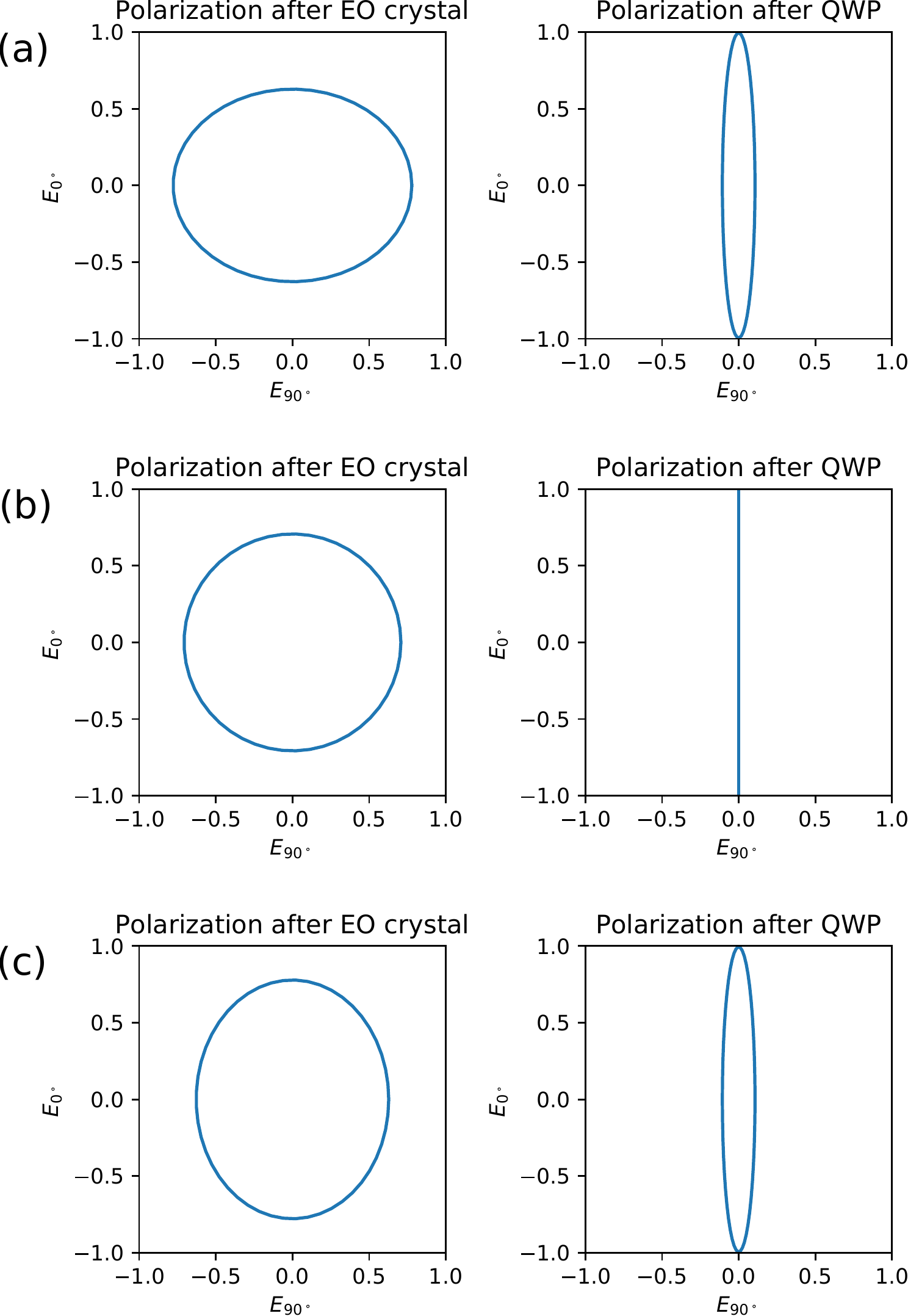}
    
    \caption{(a) Polarizations before over-rotation with THz causing a phase change in the EO crystal, with $E_{THz}\approx19~kV/cm$. (b) Polarizations with THz causing a \(\pi\)/2 phase change wave plate in the EO crystal, with $E_{THz}\approx22~kV/cm$. (c) Polarizations after over-rotation with THz causing phase change wave plate in the EO crystal, with $E_{THz}\approx25~kV/cm$.}
    \label{fig:eopovrkvpcm}
\end{figure}

Polarization simulations can further display results for greater THz electric field induced phase changes. When the phase change is \(\pi\)/2 or greater, over-rotation occurs. As seen in Figure \ref{fig:eopovrkvpcm}(b), the \(E_{0^{\circ}}\) value after the QWP has reached a maximum. However, the polarization before the QWP is circularly polarized. This means that its \(E_{0^{\circ}}\) value can continue to increase. Figures \ref{fig:eopovrkvpcm}(a) and \ref{fig:eopovrkvpcm}(c) show the polarizations before and after over-rotation, respectively. The polarizations after the QWP are identical, but the polarizations before the QWP are not. This information can be used to correct for over-rotation.

\begin{figure}[H]
    \centering
 \includegraphics[width=3in]{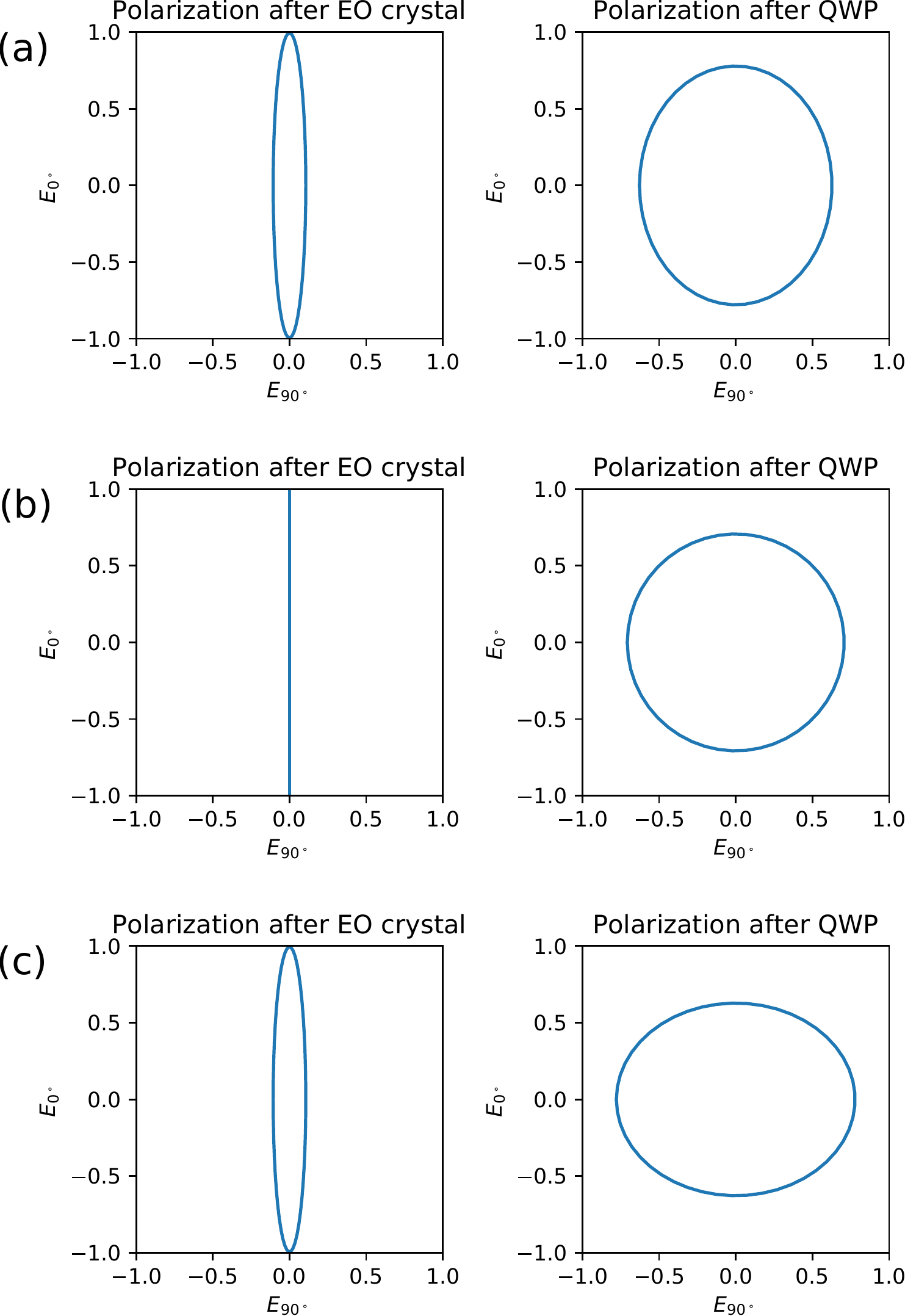} 
    
    \caption{(a) Polarizations with THz causing a phase change less than \(\pi\) in the EO crystal, with $E_{THz}\approx41~kV/cm$. (b)  Polarizations with THz causing a \(\pi\) phase change wave plate in the EO crystal, with $E_{THz}\approx44~kV/cm$. (c) Polarizations with THz causing a phase change more than \(\pi\) in the EO crystal, with $E_{THz}\approx47~kV/cm$.}
    \label{fig:eopovrpikvpcm}
\end{figure}

The polarization before the QWP will also reverse at high enough THz fields, when the phase change is at \(\pi\). As see in Figure \ref{fig:eopovrpikvpcm}(b), the \(E_{0^{\circ}}\) amplitude before the QWP has reached a maximum. However, \(E_{0^{\circ}}\) after the QWP is circularly polarized, and continues to decrease as \(\pi\) is passed, as seen with Figures \ref{fig:eopovrpikvpcm}(a) and \ref{fig:eopovrpikvpcm}(c). 

This method of going back and forth between information from before and/or after the QWP gives results of probe polarizations without adding any limit to the strength of the THz field measurable. The information could be garnered in a variety of ways. Scans could be taken with and without the QWP. Using data from different scans may not be ideal, but with automation a single scan could be done, with the QWP removed and replaced at each point in the movable stage used to map the THz field. Another option is the beam could be split before the QWP, going to another set of a Wollaston prism and photodiodes. One more option is a pulsed variable waveplate could be used in place of a QWP, that switches between no waveplate and QWP. In all cases, calibration would have to be done so a change in THz causes the same change in probe polarization intensities before and after (or with and without) the QWP. 

\section*{Discussion}

\begin{figure}[H]
\centering
\includegraphics[scale=0.65]{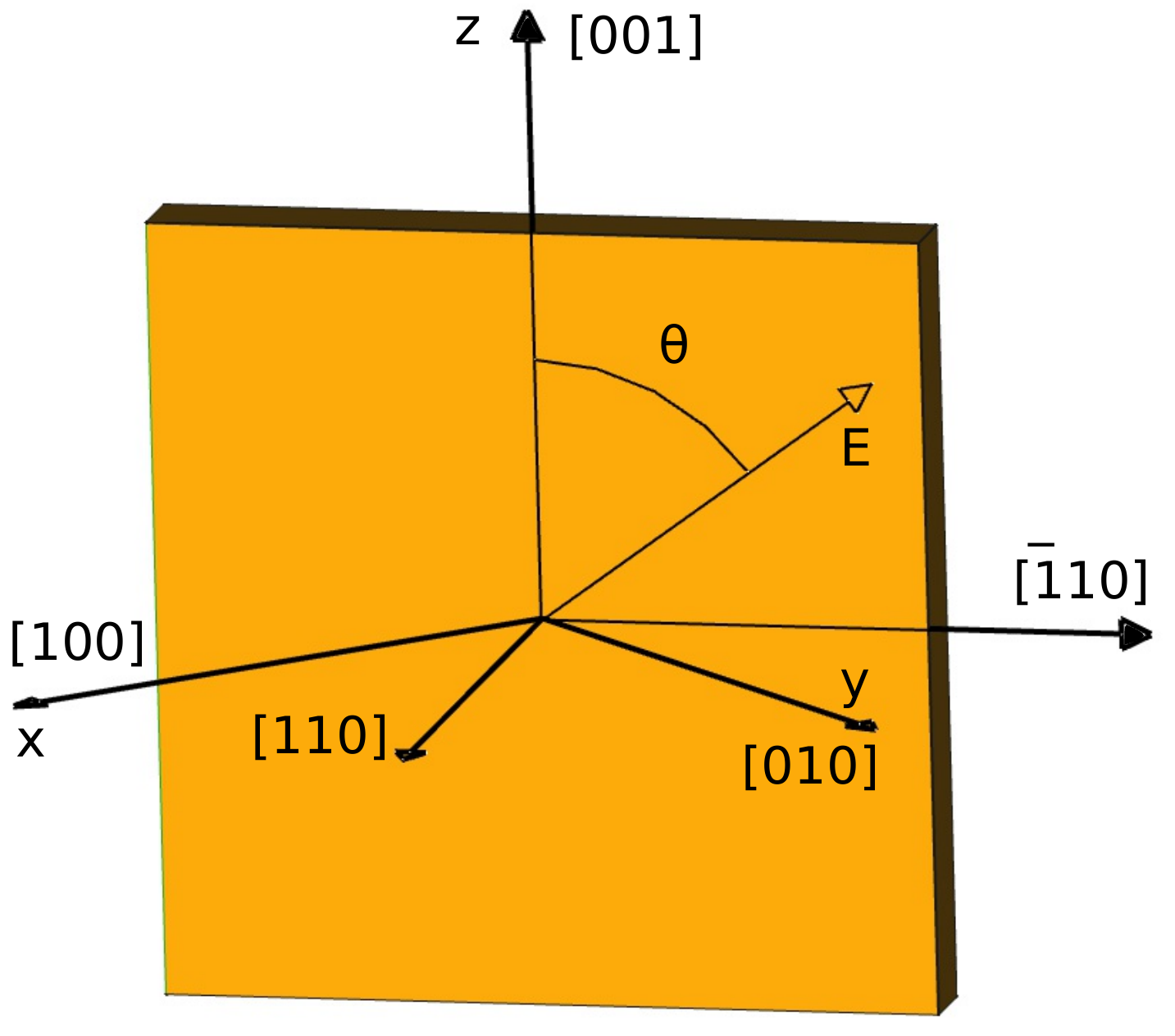}
\caption{(110) oriented ZnTe crystal with x,y, and z axes and E angle $\theta$.}
\label{znteaxes}
\end{figure}

General intensity differences in the polarizations, to lowest order in THz field, have been analyzed \cite{planken2001measurement}. Here we consider the special case, where the intensity difference is maximized to simulate the polarizations, $\theta_{THz} = \theta_{probe}= 90^{\circ}$. This is the typical configuration used in experiments and could easily be generalized to other configurations. To understand the polarizations, we first start with the index ellipsoid equation obtained from ZnTe in the presence of an electric field \cite{yariv1989quantum},
\begin{equation}
\frac{x^{2}}{n_{0}^{2}}+\frac{y^{2}}{n_{0}^{2}}+\frac{z^{2}}{n_{0}^{2}}+2r_{41}E_{Thz,x}yz+2r_{41}E_{THz,y}zx+2r_{41}E_{THz,z}xy=1
\end{equation}
where $r_{41}=3.9pm/V$ \cite{dexheimer2007terahertz} is the electro-optic coefficient of ZnTe, and $n=2.8528$ \cite{marple1964refractive} is the index of refraction of ZnTe at 800 nm. Using the coordinates as shown in Figure \ref{znteaxes}, and taking $n_{0}=n$, $E_{THz,z}=0$, and $-E_{THz,x}=E_{THz,y}=\frac{1}{\sqrt{2}}E_{THz}$, we then obtain,
\begin{equation}
\frac{x^{2}+y^{2}+z^{2}}{n^{2}}+\sqrt{2}r_{41}E_{THz}z\left ( y-x \right )=1
\end{equation}

\begin{figure}[H]
\centering
\includegraphics[scale=0.65]{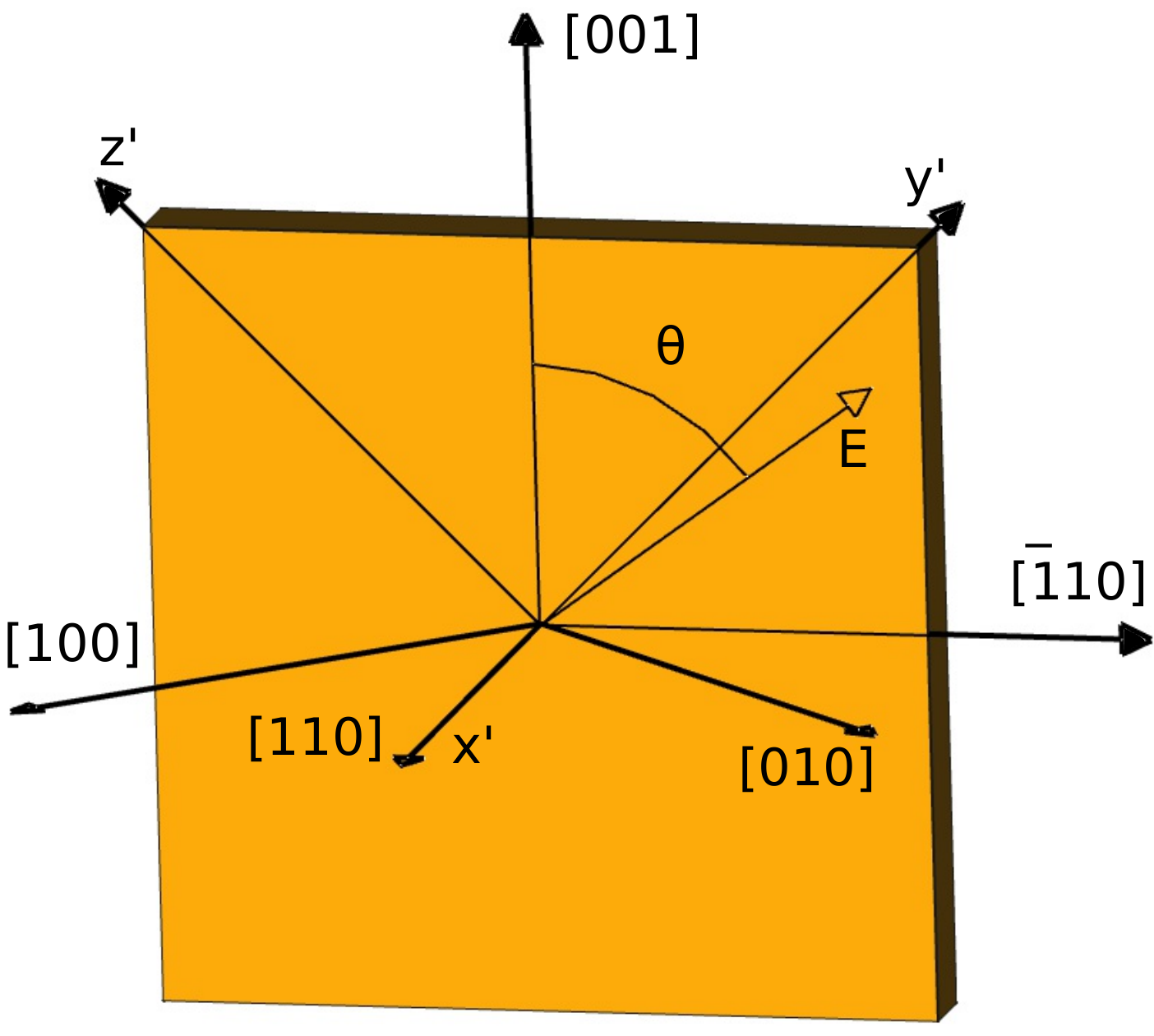}
\caption{(110) oriented ZnTe crystals with x',y', and z' axes,  and E angle $\theta$.}
\label{znteaxesprimes}
\end{figure}

To reduce the number of mixed terms, we can use a new system of axes, rotating around the z-axis $45^{\circ}$, and then a $45^{\circ}$ axis rotation around the x'-axis, with,
\begin{equation}
\begin{gathered}
x=\frac{1}{\sqrt{2}}\left ( x'-  \frac{1}{\sqrt{2}}\left ( y'-z' \right ) \right )\\
y=\frac{1}{\sqrt{2}}\left ( x'+ \frac{1}{\sqrt{2}}\left ( y'-z' \right ) \right )\\
z=\frac{1}{\sqrt{2}}\left ( y'+z' \right )
\end{gathered}
\end{equation}
which gives,
\begin{equation}
\frac{x'^{2}}{n^{2}}+y'^{2}\left (\frac{1}{n^{2}}+r_{41}E_{THz}\right )+z'^{2}\left (\frac{1}{n^{2}}-r_{41}E_{THz}\right )=1
\label{eqn:indexellipsoid}
\end{equation}
The result of these axis rotations are shown in Figure \ref{znteaxesprimes}.

When $r_{41}E_{THz}<<\frac{1}{n^{2}}$ the approximations (right expressions),
\begin{equation}
\begin{split}
n_{y'}= \frac{n}{\sqrt{1 + r_{41}n^2E_{THz}}}\approx n-\frac{n^{3}}{2}r_{41}E_{THz}\\
n_{z'}= \frac{n}{\sqrt{1 - r_{41}n^2E_{THz}}}\approx n+\frac{n^{3}}{2}r_{41}E_{THz}
\end{split}
\label{eqn:napprox}
\end{equation}
are commonly used \cite{yariv1989quantum}. However, these relations would not be valid as $E_{THz}$ gets to be a few hundred MV/cm--thus we will give the more general results and include approximations when appropriate. These directionally dependent indexes of refraction can then be put into the electric field wave equations for the probe beam components, as (with the components of the probe beams at $\theta_{E_{y'}}=45^{\circ}$ and $\theta_{E_{z'}}=-45^{\circ}$), 
\begin{equation}
E_{y'}=\sqrt{\frac{I_{p}}{2}}e^{i\left ( \omega t-\left ( \frac{\omega }{c} \right )n_{y'} L \right )}
\end{equation}
\begin{equation}
E_{z'}=\sqrt{\frac{I_{p}}{2}}e^{i\left ( \omega t-\left ( \frac{\omega }{c} \right )n_{z'} L \right )}
\end{equation}
where $\omega$ is the probe beam angular frequency, $c$ is the speed of light, $t$ is time, and $L$ is the depth of the crystal in the $x'$ direction.  

However, squaring and subtracting these field directions do not give the polarization intensity differences we are looking for, $I_{\theta=0} - I_{\theta=90^{\circ}} = \Delta I$. These intensities can be obtained from Stoke's parameter $S_{2}$ (or U), as \cite{collett2005field},
\begin{equation}
\Delta I =S_{2}=U=\left | E_{y'+45^{\circ}} \right |^{2}-\left | E_{y'-45^{\circ}} \right |^{2}
=\left \langle E_{y'}E{_{z'}}^{*} \right \rangle+\left \langle E_{z'}E{_{y'}}^{*} \right \rangle
\label{eqn:deltai}
\end{equation}
This then gives, along with using equation \eqref{eqn:napprox} for the approximations,
\begin{equation}
\Delta I_{Before~QWP}=I_{p}\cos\left (\frac{\omega L}{c} \left ( n_{y'} - n_{z'} \right ) \right ) \cong I_{p}\cos\left (\frac{\omega }{c}n^{3}r_{41}E_{THz}L  \right )
\label{eqn:intbefqwp}
\end{equation}
With the addition of a QWP at $\theta_{QWP}=45^{\circ}$, we would use 
$e^{\frac{i\pi}{4}}$
$
\begin{pmatrix}
1 & 0\\ 
0 & -i
\end{pmatrix}
$
$
\begin{pmatrix}
E_{y'}\\ 
E_{z'}
\end{pmatrix}
$
with equation \eqref{eqn:deltai} and obtain, along with the approximation \eqref{eqn:napprox},
\begin{equation}
\Delta I_{After~QWP}=I_{p}\sin\left (\frac{\omega L}{c} \left ( n_{y'} - n_{z'} \right ) \right  ) \cong I_{p}\sin\left (\frac{\omega }{c}n^{3}r_{41}E_{THz}L  \right )    
\label{eqn:intaftqwp}
\end{equation}
For a small argument of the sine, this gives the standard low THz field result, where the $\Delta I_{After}\propto E_{THz}$, which is commonly used (see equation 9 in \cite{planken2001measurement}). In general, we would have (equation 8 in \cite{planken2001measurement})
\begin{equation}
\Delta I_{After~QWP}=I_{p}\sin(2( \theta_{probe} - \phi )  \sin\left (\frac{\omega L}{c} \left ( n_{y'}(\theta_{THz}) - n_{z'}(\theta_{THz}) \right ) \right  )
\label{eqn:planken8}
\end{equation}
(of which equation \eqref{eqn:intaftqwp} here is a special case of). In the expressions above, we assumed $\varphi_{Planken~et~al.}=\theta_{probe} =90^\circ$, $\alpha_{Planken~et~al.}=\theta_{THz}=90^\circ$, and $\theta_{Planken~et~al.}=\phi=45^\circ$ (with respect to the horizontal axis, $\phi=90^{\circ}-\theta$). This is the configuration, which gives the maximum perpendicular polarization intensity change of a linearly polarized beam with a waveplate, where the waveplate fast axis would be at $45^\circ$ from the polarized beam angle. Thus the first sine would be equal to 1. The purpose of the derivation above for the special case $\theta_{THz} = \theta_{probe}= 90^{\circ}$ is to show where equation \eqref{eqn:intbefqwp} comes from and to illustrate the result of axes rotations.

The parametric plots of the electric field polarizations, are obtained by using $e^{-i\omega t}$ and the horizontally polarized Jones vector $\binom{1}{0}$. The general equations for waveplates with a fast axis at an angle $\phi$, and phase retardation $\Gamma$ are then \cite{gil1987obtainment},
\begin{equation}
e^{-\frac{i\Gamma }{2}}\begin{pmatrix}
\cos^{2}\phi +e^{i\Gamma }\sin^{2}\phi  &\left ( 1-e^{i\Gamma } \right )\cos\phi \sin\phi  \\ 
\left (1-e^{i\Gamma }  \right )\cos\phi \sin\phi  & \sin^{2}\phi +e^{i\Gamma }\cos^{2}\phi 
\end{pmatrix},
\label{eqn:waveplatematrix}
\end{equation}
where $t$ is taken over a full wavelength. The phase for the ZnTe crystal with a fast axis at $\phi=45^{\circ}$ is
\begin{equation}
\Gamma_{ZnTe}=\frac{\omega}{c}\Delta nL=\frac{\omega}{c}n^{3}r_{41}E_{THz}L 
\label{eqn:phaseznte}
\end{equation}
and a QWP at $\phi=45^{\circ}$ has a phase, 
\begin{equation}
\Gamma_{QWP}=\pi/2.
\label{eqn:qwp}
\end{equation}
For a length $L=1~mm$, over-rotation occurs at around 22~kV/cm as can be seen in Figure \ref{fig:eopovrkvpcm}. Reflection of THz on the ZnTe crystal can also be taken into account \cite{hirori2011single}, aside from also taking the detector response function into account \cite{lee2009principles}.

\begin{figure}[H]
\centering
\includegraphics[width=\textwidth]{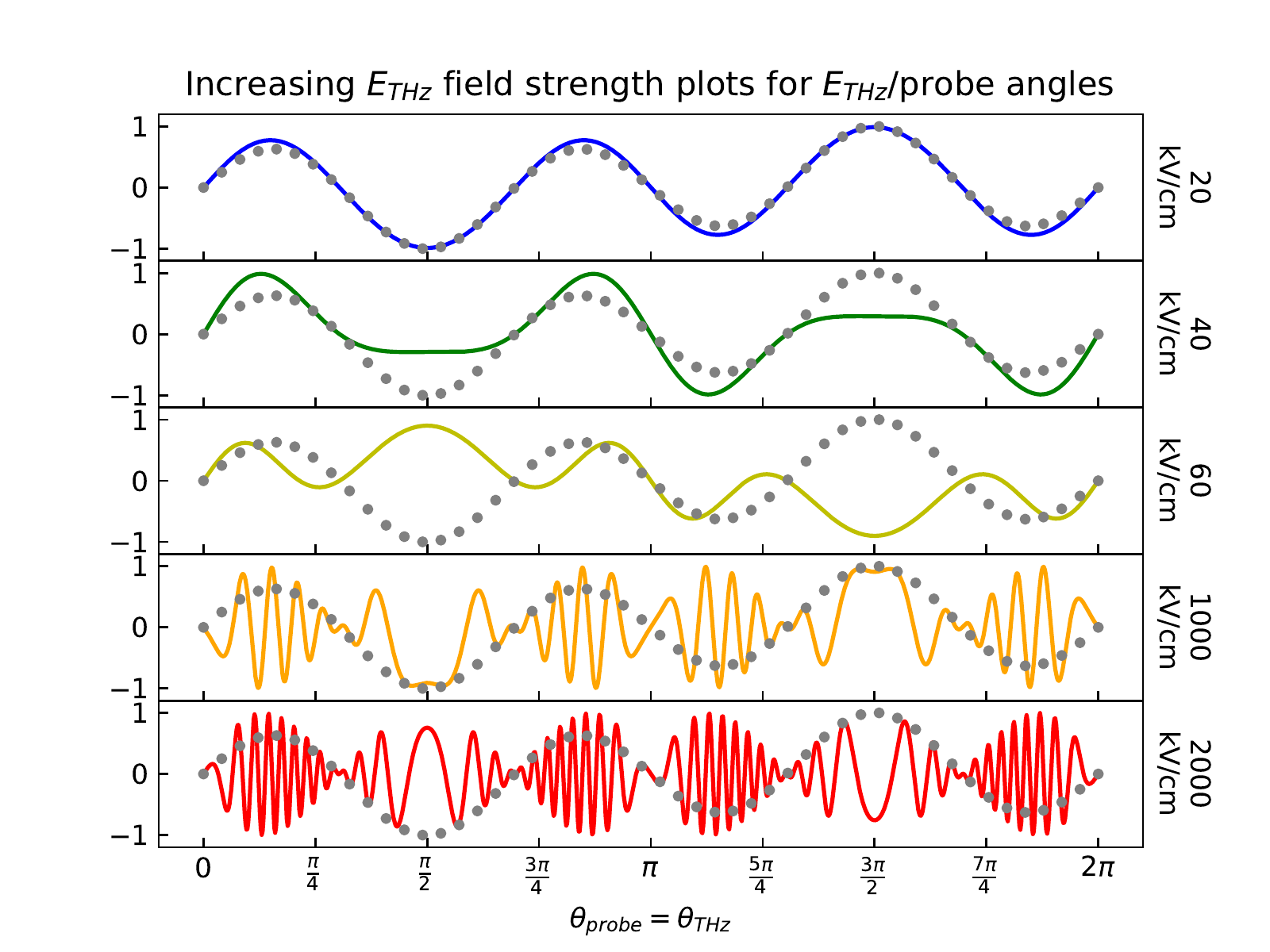}
\caption{Different magnitudes of $E_{THz}$ using equation \eqref{eqn:planken8}. The dotted lines are normalized plots using the low field approximation, where the second sine term in equation \eqref{eqn:planken8} is taken to be its argument--which would increase linearly with higher fields if it were not normalized. The vertical axis -1 to 1 values show relative peaks strengths for the given $E_{THz}$ field strengths at $\theta_{probe}$ angles. The QWP in this setup also rotates so its fast axis always $45^{\circ}$ from $\theta_{probe}$.}
\label{plankeneqn8}
\end{figure}

\begin{figure}[H]
\centering
\includegraphics[width=\textwidth]{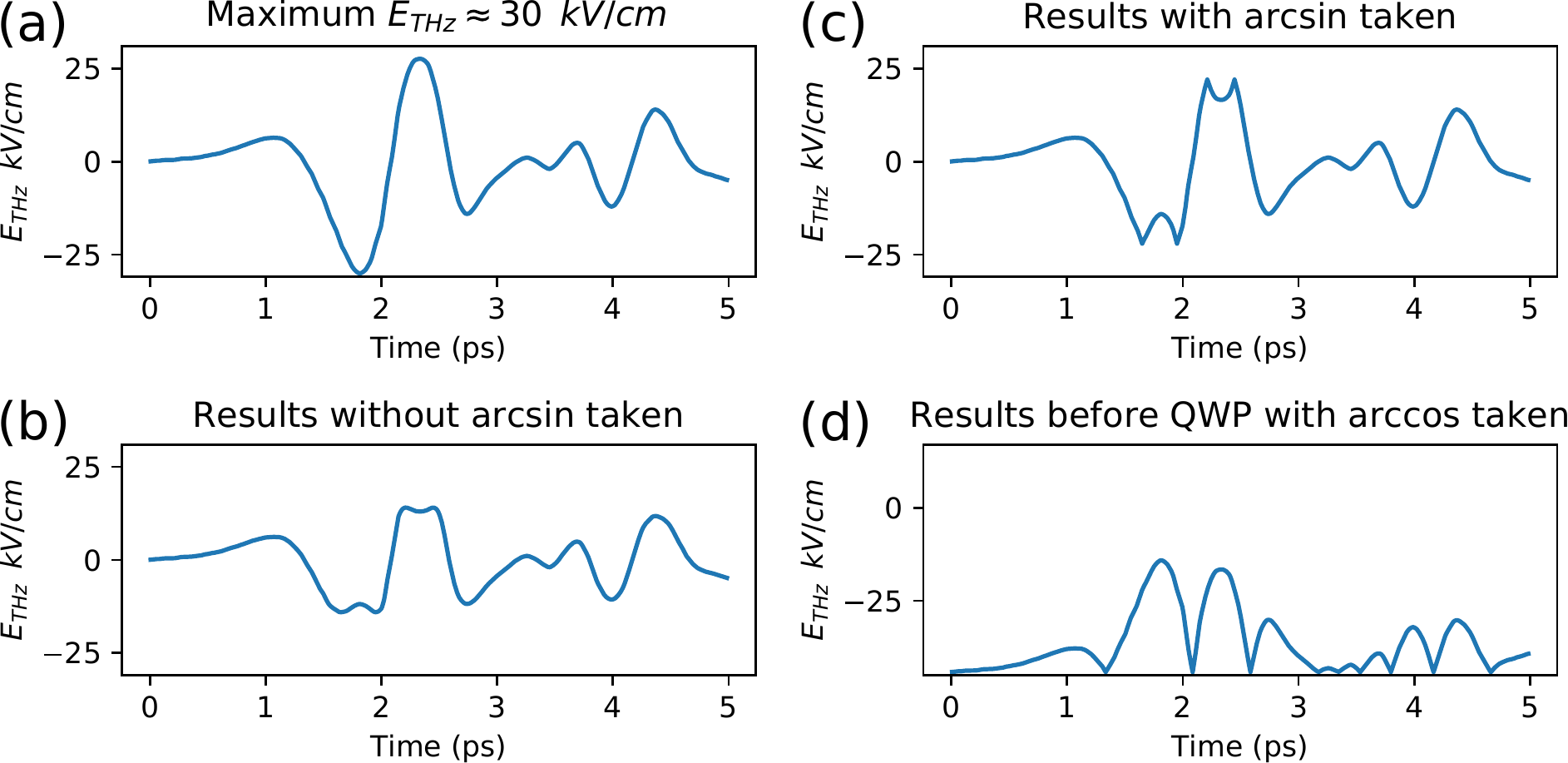}
\caption{(a) Digitized data taken from Ibrahim et al. \cite{ibrahim2016ultra}, normalized to an over-rotation occurring at $E_{THz}\approx22~kV/cm$. Results are displayed after the QWP showing over-rotation (b) without and (c) with the arcsin taken. (d) Results before the QWP are shown with the arccos taken.}
\label{ibrahimdata30kvpcm}
\end{figure}

With lower THz field, the sine in $\Delta I_{After~QWP}$ is usually taken in its first order approximation \cite{planken2001measurement}. With higher fields this would no longer be possible. Figure \ref{plankeneqn8} shows increasing $E_{THz}$ signal strengths at increasing probe and THz angles, using equation \eqref{eqn:planken8}. The angle $\theta$ of concern here occurs at $\pi/2$. The peak amplitude increases until over-rotation at around 22 kV/cm, then decreases, becoming negative at values above 44 kV/cm--where a $\pi$ phase change occurs. This would correspond to the dips in the peaks going to the negative of their initial values, as explained below and shown in Figure \ref{ibrahimdata30kvpcm}. At higher fields (1-2 MV/cm) using $E_{THz}$ horizontal (at $\pi/2$) is more likely to give accurate results, since this angle shows a smaller rate of change in the response (see Figure \ref{plankeneqn8}), though the envelope becomes more filled as the field strength continues to increase. Also, due to  $n_{x'}= \left ( \frac{1}{n^2}+ r_{41}E_{THz} \cos{\theta_{probe}} \right )^{-0.5} $ in the index ellipsoid (equation 5 in Planken et al. \cite{planken2001measurement}),
\begin{multline}
x'^{2}\left ( \frac{1}{n^2}+ r_{41}E_{THz} \cos{\theta_{probe}} \right )+\\y'^{2}\left (\frac{1}{n^{2}}+r_{41}E_{THz}(\cos\theta_{probe} \sin^{2}\phi +\cos{(\theta_{probe}+2\phi)}) \right )\\+z'^{2}\left (\frac{1}{n^{2}}-r_{41}E_{THz}(\cos\theta_{probe} \cos^{2}\phi -\cos{(\theta_{probe}+2\phi)}) \right )=1
\label{eqn:indexellipsoidplanken}
\end{multline}
(which is a general version of equation \eqref{eqn:indexellipsoid} here, with $\phi=45^{\circ}$ giving the last index ellipse rotation), higher fields as ``small" as 4 GV/cm would have an effect on phase matching as the field changes if $E_{THz}$ is not horizontal, $\theta_{THz}=90^{\circ}$. Thus having the laser probe and $E_{THz}$ horizontal to a vertical ZnTe [001] axis is the best option at high fields.

By not ignoring the full sine or cosine in either $\Delta I$, intensity differences could give accurate $E_{THz}$ values at all current ranges. Also, if over-rotation has occurred and the cosine is used with $\Delta I_{Before~QWP}$, and a phase change of $\pi$ has not been reached, this could give $E_{THz}$ in one direction for each peak. In THz science, positive and negative THz fields are generally mixed to take a Fourier Transform of the results. Although samples being examined generally have inversion symmetry in the direction of the positive and negative THz fields, this still misses an opportunity to scrutinize THz field direction interactions separately. Thus taking the Fourier Transform of THz field directions separately yields more precise information on a sample, and avoids having to mix data sets even when over-rotation occurs. Though when THz fields are high enough that more than a \(\pi\) phase change occurs, then both data sets before and after the QWP would have to be used. 

We now apply our results to data digitized from Ibrahim et al. \cite{ibrahim2016ultra} and shown in Figure \ref{ibrahimdata30kvpcm}. For better comparison we use their calculated over-rotation $E_{THz}\approx47kV/cm$ normalized to the over-rotation calculated here, at $E_{THz}\approx22kV/cm$. Thus their maximum field at $E_{THz}\approx64kV/cm$ occurs at $E_{THz}\approx30kV/cm$ here. The phases are then taken with equation \eqref{eqn:phaseznte}. Equation \eqref{eqn:waveplatematrix} for ZnTe and QWP waveplates is then used with horizontal polarization $\binom{1}{0}$ to find $\Delta I$. Finally, equations \eqref{eqn:intaftqwp} (with and without the arcsine argument) and \eqref{eqn:intbefqwp} are solved for $E_{THz}$, and the resulting data is plotted. The results before the QWP show definitive evidence that over-rotation occurs, since the dips in Figure \ref{ibrahimdata30kvpcm}(c) do not show up in Figure \ref{ibrahimdata30kvpcm}(d).  The results that leave out the arcsin show incorrect peak values, and distorted waveforms--moreso in parts closer to over-rotation. The results with the arcsin taken can then be corrected to return the original $E_{THz}$ values. 

Over-rotation can be tested by decreasing the THz field with an absorber such as silicon to avoid reverse polarizations of over-rotation. This would give clear evidence that over-rotation occurs at the dip of the peaks (as seen in Figure \ref{ibrahimdata30kvpcm}(c)) when the wafers are not used, where the polarization intensities reverse, even though the field is higher than the data yields \cite{ibrahim2016ultra}. The data at the dip in the peak could then be flipped--using values taken before the QWP and equation \eqref{eqn:intbefqwp}--turning it into a uninterrupted peak instead, recreating the original field as seen in Figure \ref{ibrahimdata30kvpcm}(a) during the same time range. If decreasing the intensities were considered a copy of the high field, just at a smaller scale, data could be adjusted accordingly. However, any absorption of a THz field has an effect on wave form, aside from just field strength, along with indexes of refraction not being uniform across the spectrum. Thus THz field interactions with the probe in the EO crystal wouldn't be the same at low and high fields. Even more, this would only be of any use in cases where a reference THz signal could be taken with absorbers that can be removed (assuming over-rotation would not occur with a sample in place, or else this method would not work in this case either), unless the absorbers occur after the sample. Of course, lowering the intensity of a THz field before a sample would defeat the purpose of having a high field THz system. Another way to avoid over rotation would be to decrease the THz peak field detection by changing the $E_{THz}$ and/or probe angles, decreasing the signal to noise ratio and dynamic range--which is generally not preferred. Trying different angle parameters with a probe polarization code can vary results under different circumstances. This would mean that equation \eqref{eqn:intaftqwp} could not be used and reverting back to the general angle case (equation \eqref{eqn:planken8}) from Planken et al. would be necessary \cite{planken2001measurement}. Thus using the maximized detection angles--with both the probe and THz horizontal--is the preferable option, while avoiding over-rotation with the methods described in this article.

As shown above, by having the complete polarization depictions of EO sampling, this enables us to understand the corrections to perform in case of over-rotation at high fields. In particular, we can use all polarization data from before and after the QWP, along with the proper calculations, to obtain an accurate reading of the THz fields even at high fields.

\bibliographystyle{unsrt}
\bibliography{EOPol}

\end{document}